\documentstyle[12pt,titlepage]{article} % Create extra title page
\title{Long Distance Contributions to B-Decays into Higher K-Resonances}
\author{Mohammad R. Ahmady, Dongsheng Liu and Zhijian Tao}          % Declares
\date{January, 1993}   % Deleting this command produces today's date.
\def\_{\rule{.3em}{.15ex}}  % Get underscore by typing \_.

\begin{titlepage}
\begin{flushright}
IC/93/26
\end{flushright}
 \begin{center}
  \vspace{0.75in}
  {\bf {\LARGE Long Distance Contributions to B-Decays into Higher
K-Resonances} \\
  \vspace{0.75in}
  Mohammad R. Ahmady, Dongsheng Liu and Zhijian Tao} \\
  International Centre For Theoretical Physics \\
  P.O. Box 586 \\
  34100 Trieste, Italy\\
  \vspace{1in}
  ABSTRACT \\
  \vspace{0.5in}
  \end{center}
  \begin{quotation}
  \noindent We use the new symmetries in the heavy quark limit to calculate the
exclusive B-decays $B \rightarrow K^i \psi $ where $K^i$ are higher resonances
of K-meson.  We then estimate the long distance contributions to $B \rightarrow
K^i \gamma $ decays via $\psi - \gamma$ conversion.  The results point to
around $30 \%$ uncertainty in the top quark mass determined by measurements of
rare B-decays $B \rightarrow K^i \gamma$.

  \end{quotation}
\end{titlepage}

\begin{document}           % End of preamble and beginning of text.
%\maketitle                 % Produces the title.
%\tableofcontents          % Must follow by a \newpage
%\newpage
%\includeonly{file1}       % Cause the argument in \include{} to be processed
%\include{file1}
%\input{your_file}

\newcommand{\da}{\mbox{$\scriptscriptstyle \dag$}}
\newcommand{\lag}{\mbox{$\cal L$}}
\newcommand{\tr}{\mbox{\rm Tr\space}}
\newcommand{\fc}{\mbox{${\widetilde F}_\pi ^2$}}
\newcommand{\ns}{\textstyle}
\newcommand{\si}{\scriptstyle}

Rare B-decays are vital testing grounds for Standard Model and therefore have
received a lot of theoretical attentions \cite{RBD,ALT,AOM}.  Besides examining
the electroweak theory at the one-loop level, and providing quantitative
informations on yet undetermined top quark mass and CKM matrix elements
$V_{td}, V_{ts}, V_{tb}$, radiative rare B-decays could also be sensitive to
the new physics beyond the Standard Model \cite{ALI}.

Our focus in this paper will be on the decays $B \rightarrow K^i \gamma$, where
$K^i$ are various K-meson excited states.  Several authors have calculated
these decays based on the dominance of the short-distance (SD) penguin operator
\cite{ALT,AOM}.  In fact, in reference \cite{AOM} the new symmetries at the
heavy quark limit \cite{HQET,IW} have been used to reduce the number of the
form factors required for the hadronic matrix elements.  This is done by
classifying the excited states of K-meson into spin-doublets, and therefore, in
the limit where {\bf s} quark is considered heavy as well as {\bf b} quark, the
six form factors parametrizing the hadronic matrix elements related to B-decays
to each spin-doublet are reduced to a single Isgur-Wise function \cite{IW}.
These functions that represent the underlying QCD dynamics of hadrons, are
obtained from some model calculations.

We are motivated by the results reported by Ali, Ohl and Mannel \cite {AOM},
which states that the SD branching fraction for the channel $K_2 ^* (1430)$ is
3 to 4 times larger than the one for $K^* (892)$.  Our purpose in this paper is
to calculate the long-distance (LD) contributions to $B \rightarrow K^i \gamma$
decays via vector meson dominance (VMD) and compare the results with the SD
contributions.  This is very important for extraction of information on, for
example, top quark mass from measurement of these decays.  The reason is as
follows: The SD contributions are dominated by penguin operator, which is
sensitive to top quark mass.  On the other hand, the LD contributions are due
to four-fermion operators, and therefore, insensitive to top quark mass.
Consequently, before drawing any conclusion on top quark mass from $B
\rightarrow K^i \gamma$ decay, we ought to know the relative magnitude of the
two channels contributing to these processes.  In fact, there have been some
model dependent estimates

f the LD contributions to $B \rightarrow K^* \gamma$, $B \rightarrow K^* \ell
^+ \ell ^-$ and $B \rightarrow K \ell ^+ \ell ^-$ \cite {DTP,DPR,GP}.

First we calculate $B \rightarrow K^i \psi$ decays using heavy quark
symmetries.  For this we follow reference \cite{AOM} in classifying different
excited states of K-meson and also in using trace formalism which was
formulated in \cite{BK,GG,KS} and extended to excited states in \cite{FALK}.
We then use $\psi - \gamma$ conversion to estimate LD contributions to $B
\rightarrow K^i \gamma$ processes.

The effective Hamiltonian relevant to $B \rightarrow K^i \psi$ processes can be
written as \cite{DTP}:
\begin{equation}
 H _{eff} = C  f_{\psi} \bar s \gamma _{\mu} (1- \gamma _5 )b \epsilon ^{\mu}
_{\psi}.
\end{equation}
 where $$ f_{\psi} \epsilon ^{\mu} _{\psi} = <0| \bar c \gamma ^{\mu} c| \psi >
,$$ \\
\begin{equation}
 C= {{G _F} \over {\sqrt 2}} (c _1 + c _2 /3 ) V _{cs} ^* V _{cb}.
\end{equation}
$c_1$ and $c_2$ are QCD improved Wilson coefficients:
\begin{equation}
{\displaystyle
\begin{array}{l}
\displaystyle{c_1 = \frac {1}{2} \left ( {\left [ \frac {\alpha _s ( \mu
)}{\alpha _s ( m _W)} \right ] }^{-6/23} - {\left [ \frac {\alpha _s ( \mu
)}{\alpha _s ( m _W)} \right ] }^{12/23} \right )}, \\
\displaystyle{c_2 = \frac {1}{2} \left ( {\left [ \frac {\alpha _s ( \mu
)}{\alpha _s ( m _W)} \right ] }^{-6/23} + {\left [ \frac {\alpha _s ( \mu
)}{\alpha _s ( m _W)} \right ] }^{12/23} \right )}.
\end{array}
}
\end{equation}
In the limit where {\bf b} and {\bf s} quarks are considered heavy, matrix
elements of (1) are calculated by taking a trace:

\begin{equation}
\begin{array}{l}
<K^i (v^{\prime} )| \bar s \gamma _{\mu} (1- \gamma _5 )b |B(v)> = \\
Tr \left [\bar \Re ^i ( v^{\prime}) \gamma _ \mu (1- \gamma _5 ) \Re (v) M (v,
v^{\prime}) \right ].
\end{array}
\end{equation}
$\Re ^i ( v^{\prime})$ and $\Re  ( v)$ are the matrix representations of $K^i$
and $B$ respectively, $v$ and $v^{\prime}$ are velocities of initial and final
state mesons \cite{FALK}.  $M$ which represents the light degrees of freedom is
related to Isgur-Wise functions.  In fact, heavy quark symmetries have been
applied to some exclusive $B \rightarrow (K, K^*)+( \psi , \psi ^{\prime})$
decays, in which the universal function has been determined by the best fit to
semileptonic decay data \cite {AL}.  We follow the notations of reference
\cite{AOM} in labeling the Isgur-Wise functions relevant to B-decays to each
spin-doublet of K-meson excited states.  With this in mind, we arrive at the
following expression for the decay rates:
\begin{equation}
\begin{array}{l}
\Gamma (B \rightarrow K^i \psi ) = \displaystyle{\frac {C^2 f^2 _{\psi}}{16 \pi
} g(m _B , m _{K^i} , m _{\psi} )   m _{K^i} F_i (x _i) {\vert \xi _I (x _i)
\vert }^2 },\\
\displaystyle{g(m _B , m _{K^i} , m _{\psi} ) =  {\left [ {\left (1- \frac {m^2
_{\psi} }{m^2 _B} -\frac {m^2 _{K^i}}{m^2 _B} \right )}^2 - \frac {4 m^2 _{K^i}
m^2 _{\psi} }{m^4 _B} \right ]}^{1/2}} ,\\
\displaystyle{x _i = v. v ^{\prime} = \frac {m^2 _B + m^2 _{K^i} - m^2 _{\psi}
}{2 m _B m _{Ki} }} .
\end{array}
\end{equation}
$\xi _I (x) , I= C,E,F,G$ are the Isgur-Wise functions for various
spin-doublets .  We will find it useful, for our formulation of $\psi - \gamma$
conversion, to separate the total decay rate (5) into transverse $\Gamma ^T (B
\rightarrow K^i \psi )$ and longitudinal $\Gamma ^L (B \rightarrow K^i \psi )$
parts, corresponding to transverse and longitudinally polarized $\psi$ in the
final state respectively.  Consequently, the functions $F_i  (x)$ for various
excited states of K-meson are decomposed into transverse $F^T _i (x)$ and
longitudinal $F_i ^L (x)$ contributions.  In table 1, we tabulate these
functions for various K-meson excited states.

The vector meson dominance approximation for $A \rightarrow B \gamma$
transition can be written as:
\begin{equation}
< \gamma B | T | A > = \sum _V  \frac {(e _Q e f_V )}{m^2 _V} <VB|T|A>,
\end{equation}
where $f_V$ is the decay constant of vector meson $V$, and $e_Q$ is the charge
of the constituent quark.  In fact, VMD implies the replacement of
electromagnetic current with vector mesons.  The conservation of the
electromagnetic current  requires that the polarization of the intermediate
vector meson should satisfy $q. \epsilon _{\psi} =0$ condition, where $q$ is
the four-momentum of the vector meson (which is the same as the four-momentum
of the final state photon) \cite {BOOK}.  This is true automatically when the
vector meson is on mass shell.  As a result, in our scheme, we choose to sum
over only those polarizations of the intermediate vector mesons that satisfy
the above condition.  This scheme guarantees the gauge invariance of the matrix
elements, hence we can use the specific gauge, i.e., the Coulomb gauge, to do
the calculations.  For $B \rightarrow K^i \gamma$ decays $V$ stands for $\psi ,
\psi ^{\prime} , \psi ^{\prime \prime}$, etc., as other contributions are
suppressed due to CKM mixing ang

es.  These radial excitations of $\psi$ all contribute coherently leading to:
\begin{equation}
\Gamma ^{LD} (B \rightarrow K^i \gamma ) = {\it F} \Gamma ^T (B \rightarrow K^i
\psi ) \vert _{m_{\psi} ^2 =0},
\end{equation}
where
\begin{equation}
{\displaystyle
{\it F} = \frac {4 \pi \alpha e_Q ^2}{f_{\psi} ^2} {\left ( \sum _{V = \psi ,
\psi ^{\prime} , \psi ^{\prime \prime} ...} \frac {f_V ^2}{m^2 _V} \right )
}^2,
}
\end{equation}
and $\Gamma ^T (B \rightarrow K^i \psi )$ is the transverse part of the decay
$B \rightarrow K^i \psi $.

$f_V$ for various radial excitations of $\psi$ can be obtained from
experimental data on $\Gamma (V \rightarrow e^+ e^- )$, for example , $f^2
_{\psi} = 2.38 \pm 0.14  {(GeV)}^4$ by comparison between theoretical decay
rate for $\psi \rightarrow e^+ e^-$ (including order $\alpha _s $ QCD
corrections) and $\Gamma ( \psi \rightarrow e^+ e^- ) = 4.7 \pm 0.3 $KeV.
Substituting numerical values for $f_V, m_V, $ and $e_Q = 2/3$ in (8) we obtain
\cite {DTP}:
\begin{equation}
{\it F} = 4.804 \times {10}^{-3}.
\end{equation}

On the other hand, using (1) we can obtain the inclusive decay rate $\Gamma (B
\rightarrow X_s \psi )$ which we equate to $\Gamma (b \rightarrow s \psi ) $:
\begin{equation}
\Gamma (b \rightarrow s \psi )  = \displaystyle{\frac {C^2 f^2 _{\psi} }{8 \pi
m_b m^2 _{\psi}} g(m _b , m _s , m _{\psi} ) [m^2 _b (m^2 _b + m^2 _{\psi}) -
m^2 _s (2 m^2 _b - m^2 _{\psi} ) + m^4 _s -2 m^4 _{\psi}]},
\end{equation}
where $\psi$ in this decay is directly produced.  Consequently, using \cite
{DTP}
\begin{equation}
\frac {\Gamma (b \rightarrow s \psi )}{\Gamma (b \rightarrow all )} = (1.0 \pm
0.24) \times {10}^{-2},
\end{equation}
and (5), (7) and (9) we obtain the branching ratios for LD processes:
\begin{equation}
{\displaystyle
\begin{array}{l}
BR ^{LD} (B \rightarrow K^i \gamma ) = \displaystyle{\frac {\Gamma ^{LD} (B
\rightarrow K^i \gamma )}{\Gamma (b \rightarrow all )}} = 2.402 \times
{10}^{-5}\\
\displaystyle{ \times \frac {g(m_B,m_{K^i},0)m_{K^i}m_b m_{\psi}
^2}{g(m_b,m_s,m_{\psi})[m_b ^2 (m_b ^2 +m^2 _{\psi} )-m_s ^2 (2m_b ^2 - m^2
_{\psi})+m^4 _s -2m^4 _{\psi}]} F_i ^T ( x^{\circ} _i) {\vert \xi _I (x
^{\circ} _i) \vert}^2},
\end{array}
}
\end{equation}
where
$${x ^ {\circ} _i} = \displaystyle{\frac {m_B ^2 + m^2 _{K^i}}{2 m_B
m_{K^i}}}.$$  We use the following numerical values (in GeV) for mass
parameters in our calculations:
$$\begin{array}{lll}
m_b =4.9  &  m_s =0.15 \\
m_{\psi} = 3.10  &  m_B = 5.28
\end{array}
$$

In reference \cite {AOM} the Isgur-Wise functions $\xi _I (x_{\circ}),
I=C,E,F,G$ have been obtained using the wavefunction model of Isgur, Scora,
Grinstein and Wise \cite {ISGW}.  Inserting these functions in (12) leads to
our results for branching ratios which are presented along with R, the ratio
between long and short distance amplitudes, in table 2.  We should point out
that due to the cancellation of the Isgur-Wise function, appearing in both long
and short range contributions, the ratio R is model independent.  Also, the
$m_{K^i}$ dependent part of R is of the form $\frac {m_B}{\sqrt {m_B ^2
-m_{K^i} ^2}}$, leading to small change in the value of R for various
K-resonances.  This implies that the LD branching ratios follow the same
pattern as the SD one reported in reference \cite {AOM}.

In conclusion, as reflected in table 1, the transverse part of $0^-$ to $0^-$
transitions vanishes.  This, in our gauge invariant scheme, results in null
long-range contributions to $0^- \rightarrow 0^- + \gamma$ decays, which is
consistent with general helicity argument.

We observe from table 2, the long-distance contributions are about $(18-21) \%$
of the penguin contributions to $B \rightarrow K^i \gamma$ decays.  As
mentioned earlier, the knowledge of this ratio is important in determination of
top quark mass from rare B-decays.  The total amplitude for $B \rightarrow K^i
\gamma$ can be written as:
\begin{equation}
A^{total} =(1 \pm R) A^P ,
\end{equation}
where $A^P$ is the amplitude due to penguin operator \cite {GP}.  In (13) we
have assumed real phase between short and long range contributions, hence
neglecting CP noninvariance.  Taking into account the $m_t$ dependence of
$A^P$, we reach the conclusion that $m_t$ can be determined up to a factor $(1
\pm 1.7R)$.  Therefore, our results based on leading heavy quark symmetries
indicate that there is at least $30 \%$ uncertainty in the determination of top
quark mass from rare B-decays $B \rightarrow K^i \gamma$.

Finally, the gauge invariance issue in calculating vector meson dominant decays
requires careful attention, as the effective Hamiltonian (1), when $\epsilon
_{\psi}$ is replaced by photon polarization is not gauge invariant.  For
example, in reference \cite {DTP}, the quark model calculations result in the
equality of $A_1$ and $A_2$ form factors, which in turn leads to gauge
invariant results.  In the heavy quark limit, this model-dependent equality is
not valid.  On the other hand, the modified gauge invariant effective
Hamiltonian suggested at the end of that paper, yields null long-range
contribution to $B \rightarrow K^* \gamma$ decay.  However, our scheme
guarantees the gauge invariance and at the same time leads to nontrivial LD
contributions to $B \rightarrow K^i \gamma$.

\noindent
{\bf Acknowledgement}

\noindent
The authors thank A. Ali and R. R. Mendel for useful discussions.

\newpage

\hskip -1cm
{\small
\begin{tabular}{c|c|c|c}
\hline
\multicolumn{1}{|c|}{$K^i$ Name} &
\multicolumn{1}{c|}{$ J^P $} &
\multicolumn{1}{c|}{$F_i ^L (x)$} &
\multicolumn{1}{c|}{$F_i ^T$} \cr
\hline
\multicolumn{1}{|c|}{$K (498)$} &
\multicolumn{1}{c|}{$0^-$} &
\multicolumn{1}{c|}{$(x+1)[(x+1){(m_B -m_K)}^2/m_{\psi} ^2-2]$} &
\multicolumn{1}{c|}{$0$} \cr
\hline
\multicolumn{1}{|c|}{$K^* (892)$} &
\multicolumn{1}{c|}{$1^-$} &
\multicolumn{1}{c|}{$(x+1)[(x-1){(m_B +m_{K^*})}^2/m_{\psi} ^2+2]$} &
\multicolumn{1}{c|}{$4x(x+1)$} \cr
\hline
\multicolumn{1}{|c|}{$K^* (1430)$} &
\multicolumn{1}{c|}{$0^+$} &
\multicolumn{1}{c|}{$(x-1)[(x-1){(m_B +m_{K^*})}^2/m_{\psi} ^2+2]$} &
\multicolumn{1}{c|}{$0$} \cr
\hline
\multicolumn{1}{|c|}{$K _1 (1270)$} &
\multicolumn{1}{c|}{$1^+$} &
\multicolumn{1}{c|}{$(x-1)[(x+1){(m_B -m_{K_1})}^2/m_{\psi} ^2-2]$} &
\multicolumn{1}{c|}{$4x(x-1)$} \cr
\hline
\multicolumn{1}{|c|}{$K_1 (1400)$} &
\multicolumn{1}{c|}{$1^+$} &
\multicolumn{1}{c|}{$2/3(x-1){(x+1)}^2[(x+1){(m_B -m_{K_1})}^2/m_{\psi} ^2-2]$}
&
\multicolumn{1}{c|}{$2/3x(x-1){(x+1)}^2$} \cr
\hline
\multicolumn{1}{|c|}{$K^* _2 (1430)$} &
\multicolumn{1}{c|}{$2^+$} &
\multicolumn{1}{c|}{$2/3(x-1){(x+1)}^2[(x-1){(m_B +m_{K^* _2})}^2/m_{\psi}
^2+2]$} &
\multicolumn{1}{c|}{$2x(x-1){(x+1)}^2$} \cr
\hline
\multicolumn{1}{|c|}{$K^* (1680)$} &
\multicolumn{1}{c|}{$1^-$} &
\multicolumn{1}{c|}{$2/3(x^2-1)[{(x-1)}^2{(m_B +m_{K^*})}^2/m_{\psi} ^2-2]$} &
\multicolumn{1}{c|}{$2/3x(x+1){(x-1)}^2$} \cr
\hline
\multicolumn{1}{|c|}{$K_2 (1580)$} &
\multicolumn{1}{c|}{$2^-$} &
\multicolumn{1}{c|}{$2/3(x+1){(x-1)}^2[(x+1){(m_B -m_{K_2})}^2/m_{\psi} ^2-2]$}
&
\multicolumn{1}{c|}{$2x(x+1){(x-1)}^2$} \cr
\hline
\multicolumn{1}{|c|}{$K(1460)$} &
\multicolumn{1}{c|}{$0^-$} &
\multicolumn{1}{c|}{$(x+1)[(x+1){(m_B -m_K)}^2/m_{\psi} ^2-2]$} &
\multicolumn{1}{c|}{$0$} \cr
\hline
\multicolumn{1}{|c|}{$K^* (1410)$} &
\multicolumn{1}{c|}{$1^-$} &
\multicolumn{1}{c|}{$(x+1)[(x-1){(m_B +m_{K^*})}^2/m_{\psi} ^2+2]$} &
\multicolumn{1}{c|}{$4x(x+1)$} \cr
\hline
\end{tabular}
\hskip 1cm
\center{Table 1. $F^L _i (x)$ and $F_i ^T (x)$ for various K-meson excited
states. }}

\vskip 2cm
\hskip -2.5cm
{\small
\begin{tabular}{c|c|c|c|c|c|c}
\hline
\multicolumn{1}{|c|}{$K^i$ Name} &
\multicolumn{1}{c|}{$ J^P $} &
\multicolumn{1}{c|}{Mass (MeV)} &
\multicolumn{1}{c|}{$ x _{\circ}$} &
\multicolumn{1}{c|}{$\xi _I (x_{\circ})$} &
\multicolumn{1}{c|}{$\begin{array}{l} BR(B \rightarrow K^i \gamma ) \times
{10}^5 \\  Long-distance \end{array}$} &
\multicolumn{1}{c|}{$ R =  {\left [ \frac {BR^{LD}(B \rightarrow K^i \gamma
)}{BR^{SD}(B \rightarrow K^i \gamma )} \right ]}^{1/2}$} \cr
\hline
\multicolumn{1}{|c|}{$K $} &
\multicolumn{1}{c|}{$0^-$} &
\multicolumn{1}{c|}{$497.67 \pm 0.03$} &
\multicolumn{1}{c|}{$5.346$} &
\multicolumn{1}{c|}{$0.125-0.239$} &
\multicolumn{1}{c|}{$forbidden$} &
\multicolumn{1}{c|}{$-$} \cr
\hline
\multicolumn{1}{|c|}{$K^* (892)$} &
\multicolumn{1}{c|}{$1^-$} &
\multicolumn{1}{c|}{$896.1 \pm 0.3$} &
\multicolumn{1}{c|}{$3.030$} &
\multicolumn{1}{c|}{$0.136-0.253$} &
\multicolumn{1}{c|}{$0.046 - 0.160$} &
\multicolumn{1}{c|}{$0.18$} \cr
\hline
\multicolumn{1}{|c|}{$K^* (1430)$} &
\multicolumn{1}{c|}{$0^+$} &
\multicolumn{1}{c|}{$1429 \pm 7$} &
\multicolumn{1}{c|}{$1.982$} &
\multicolumn{1}{c|}{$0.309-0.453$} &
\multicolumn{1}{c|}{$forbidden$} &
\multicolumn{1}{c|}{$-$} \cr
\hline
\multicolumn{1}{|c|}{$K _1 (1270)$} &
\multicolumn{1}{c|}{$1^+$} &
\multicolumn{1}{c|}{$1270 \pm 10$} &
\multicolumn{1}{c|}{$2.198$} &
\multicolumn{1}{c|}{$0.296-0.441$} &
\multicolumn{1}{c|}{$0.065 - 0.144$} &
\multicolumn{1}{c|}{$0.19$} \cr
\hline
\multicolumn{1}{|c|}{$K_1 (1400)$} &
\multicolumn{1}{c|}{$1^+$} &
\multicolumn{1}{c|}{$1402 \pm 7$} &
\multicolumn{1}{c|}{$2.015$} &
\multicolumn{1}{c|}{$0.307-0.451$} &
\multicolumn{1}{c|}{$0.089 - 0.193$} &
\multicolumn{1}{c|}{$0.19$} \cr
\hline
\multicolumn{1}{|c|}{$K^* _2 (1430)$} &
\multicolumn{1}{c|}{$2^+$} &
\multicolumn{1}{c|}{$1425.4 \pm 1.3$} &
\multicolumn{1}{c|}{$1.987$} &
\multicolumn{1}{c|}{$0.309-0.452$} &
\multicolumn{1}{c|}{$0.259 - 0.554$} &
\multicolumn{1}{c|}{$0.19$} \cr
\hline
\multicolumn{1}{|c|}{$K^* (1680)$} &
\multicolumn{1}{c|}{$1^-$} &
\multicolumn{1}{c|}{$1714 \pm 20$} &
\multicolumn{1}{c|}{$1.702$} &
\multicolumn{1}{c|}{$0.359-0.420$} &
\multicolumn{1}{c|}{$0.018 - 0.024$} &
\multicolumn{1}{c|}{$0.21$} \cr
\hline
\multicolumn{1}{|c|}{$K_2 (1580)$} &
\multicolumn{1}{c|}{$2^-$} &
\multicolumn{1}{c|}{$\approx 1580$} &
\multicolumn{1}{c|}{$1.820$} &
\multicolumn{1}{c|}{$0.350-0.417$} &
\multicolumn{1}{c|}{$0.071 - 0.101$} &
\multicolumn{1}{c|}{$0.20$} \cr
\hline
\multicolumn{1}{|c|}{$K(1460)$} &
\multicolumn{1}{c|}{$0^-$} &
\multicolumn{1}{c|}{$\approx 1460$} &
\multicolumn{1}{c|}{$1.946$} &
\multicolumn{1}{c|}{$-0.242 - -0.293$} &
\multicolumn{1}{c|}{$forbidden$} &
\multicolumn{1}{c|}{$-$} \cr
\hline
\multicolumn{1}{|c|}{$K^* (1410)$} &
\multicolumn{1}{c|}{$1^-$} &
\multicolumn{1}{c|}{$1412 \pm 12$} &
\multicolumn{1}{c|}{$2.005$} &
\multicolumn{1}{c|}{$-0.240 - -0.292$} &
\multicolumn{1}{c|}{$0.107 - 0.158$} &
\multicolumn{1}{c|}{$0.19$} \cr
\hline
\end{tabular}
\hskip 2.5cm
\center{Table 2. Comparison between branching ratios of long-distance and
short-distance contributions to B-decays into various K-meson resonances. }}

\end{document}